\documentclass[12pt,a4paper]{article}
\usepackage{a4wide}
\usepackage{latexsym}
\usepackage{epsf}
\usepackage{amssymb}
\begin{document}
\def\be{\begin{equation}}
\def\ee{\end{equation}}
\def\bea{\begin{eqnarray}}
\def\eea{\end{eqnarray}}

\def\pd{\partial}
\def\a{\alpha}
\def\b{\beta}
\def\g{\gamma}
\def\d{\delta}
\def\m{\mu}
\def\n{\nu}
\def\t{\tau}
\def\l{\lambda}

\def\s{\sigma}
\def\e{\epsilon}
\def\scri{\mathcal{J}}
\def\cM{\mathcal{M}}
\def\tcM{\tilde{\mathcal{M}}}
\def\RR{\mathbb{R}}

\hyphenation{re-pa-ra-me-tri-za-tion}
\hyphenation{trans-for-ma-tions}


\begin{flushright}
IFT-UAM/CSIC-99-45\\
hep-th/9911215\\
\end{flushright}

\vspace{1cm}

\begin{center}

{\bf\Large The Renormalization Group Approach to the Confining String }

\vspace{.5cm}
 
{\bf Enrique \'Alvarez\ddag}
\footnote{E-mail: {\tt enrique.alvarez@uam.es,@cern.ch}}
and {\bf C\'esar G\'omez\dag}
\footnote{E-mail: {\tt cesar.gomez@uam.es}}
\vspace{.3cm}

\vskip 0.4cm

{\it \ddag Theory Division, CERN,1211 Geneva 23, Switzerland,\\
\dag\ddag Instituto de F\'{\i}sica Te\'orica, C-XVI,
\footnote{Unidad de Investigaci\'on Asociada
  al Centro de F\'{\i}sica Miguel Catal\'an (C.S.I.C.)}
and  Departamento de F\'{\i}sica Te\'orica, C-XI,\\
  Universidad Aut\'onoma de Madrid 
  E-28049-Madrid, Spain }

\vskip 0.2cm

\vskip 1cm


{\bf Abstract}

\end{center}

\begin{quote}
The renormalization group approach towards the string representation 
of non abelian gauge theories
translates, in terms of the string sigma model beta function equations, 
the renormalization group evolution
of the gauge coupling constant and Zamolodchikov`s $c$ function. Tachyon
stability, glueball mass gap, renormalization group 
evolution of the $c$ function and the area law for
the Wilson loop are studied for
a critical bosonic string vacuum corresponding to a non abelian gauge
 theory in four dimensional space-time.  
We prove that the same intrinsic geometry for the string vacuum is universal
in some sense, reproducing the Yang-Mills
beta function to arbitrary loop order in perturbation theory.

\end{quote}


\newpage

\setcounter{page}{1}
\setcounter{footnote}{1}
\tableofcontents
\newpage

\vspace{1cm}
\section{Introduction}

Phenomenologically the string approach to non abelian gauge theories
is based on interpreting the color flux tubes responsible for
quark confinement as strings ending on  quark-antiquark pairs. From
a more theoretical point of view a string representation of a gauge
theory amounts to define vacuum expectation values
for Wilson loops as sums over random surfaces with boundary given by the loop.
After the discovery of D-branes a new way to associate strings 
with pure gauge theories has been suggested based on strong-weak coupling
duality. In that approach, based on Maldacena's conjecture \cite{maldacena},
 the strong
't Hooft's coupling regime is in correspondence with weakly coupled type
IIB string theory. The weakly coupled Yang Mills is in this case
described by strongly coupled string theory and therefore out of 
present technology. A similar problem arises when the Type IIB description
is extended to non supersymmetric theories \cite{witten} 
with the supergravity regime
being defined by 'tHooft's coupling much larger that the natural cutoff,
implying the presence of undesired degrees of freedom.

\par

In the same vein (of working out a string description of
the low energy world volume theory on D-branes) type 0 string theories
were also considered \cite{klebanov}. There we have a closed string tachyon
which means, from the point of view of anti de Sitter space-time, that we 
should work in a regime where the curvature is of the order of the string scale
i.e out of the gravity approximation . Using the vacuum expectation 
value for the tachyon
as source of the dilaton,  an approximate solution with a 
running of the said dilaton was obtained 
which is consistent
with asymptotic freedom .The
curvature in the ultraviolet (UV) region being infinity  again 
indicates that
the gravity description is not valid in the UV asymptotically free regime.
Both in the type 0 case or in type IIB with supersymmetry broken by
boundary conditions on the internal Kaluza Klein manifold the starting point
was the geometry as defined, in the near horizon limit, by a stack of D branes.

\par

A different approach, presented in a previous paper \cite{alvarezgomez},
see also \cite{akhmedov}, 
stems from the idea of using
strings to model out  the renormalization group equations of the quantum
field theory. This in particular leads to associate to
the coupling constants of the quantum field theory, some background fields
of a closed string theory and to demand that the sigma model beta functions
characterizing the string background precisely coincide with the
renormalization group equations of the non abelian gauge theory. 
In this sense the quantum field theory
beta functions appear as the imput in the string beta function equations. The
output, that is purely stringy, will consist in the specific space-time 
geometry
that will be formally defined on the space-time extended by an extra
(renormalization group) dimension. This string geometry is the one we can
use to define representations of the vacuum expectation value of the Wilson
loop in terms of random surfaces, most likely with zig-zag type of
boundary conditions \cite{polyakov}. 
Notice that the difference between the D-brane approach 
and the renormalization group approach (RGA) lies mainly
 in the way the field theory
input is introduced. In the D-brane approach this is done 
 in terms of the low energy 
physics on the 
D-brane. In the RGA approach we use instead 
the quantum field theory beta functions
to fix the dilaton behavior and employ the string beta function equations
to derive the stringy space-time geometry.

\par

The physical meaning, from the quantum
field theory point of view, of the stringy space-time geometry is related
to the renormalization group evolution of Zamolodchikov's $c$-function. More
precisely the curvature on the renormalization group dimension
will induce a focusing phenomena with the interpretation of decreasing
$c$-function \cite{zamolodchikov}. 
In summary the RGA intends to translate in pure gravitational
terms, codified in the string beta function equations, the renormalization
group evolution of both the gauge coupling constant and Zamolodchikov's 
$c$-function. The gravitational description of renormalization group
evolution is our way to understand the meaning of holography\cite{akhmedov}.

\par

We want to continue in this work the study of a critical bosonic string
background with the dilaton field defined by an asymptotically free beta
function of the type of pure ${\cal N} = 0 $ Yang-Mills. 
\par
The solution
presented in reference \cite{alvarezgomez} is generalized
in this paper to the case of
two loop Yang Mills beta function. The resulting confining metric posseses
the same intrinsic geometry that for the one loop case.
Moreover the confining geometry is universal for arbitrary
order in perturbation theory. This
confining string metric is an exact solution to the critical bosonic string
sigma model beta function equations for vanishing vacuum expectation value
of the tachyon. 
\par 
Next we consider the problem of stability of our solution 
showing that on the reduced functional space defined by demanding conservation
of energy, the linearized small fluctuation problem does not admit normalizable
solutions with negative eigenvalue. We also study the existence of a mass gap
for glueball equations as well as the area law behavior for the vacuum 
expectation value of the Wilson loop. In addition we present a careful
analysis of the geometry of the confining string vacuum with the
expected behavior of a geometric $c$-function defined in terms of the
expansion for congruences of null geodesics.

\section{Gauge Geometry}
Our starting point is the RG evolution for an asymptotically free
pure gauge theory. For simplicity we will first consider the case of the
one loop beta function equation:
\be
{\mu}\frac{g}{d{\mu}}= -  \frac{\b_0}{(4\pi)^2} g^{3}
\ee\label{beta}
for $\b$ some numerical constant depending on the gauge group and $g$ 
the Yang Mills coupling constant. (The free abelian theory can also be put
in this form, with $\b_0=0$). Next we use the standard relation
between the gauge coupling constant and the dilaton field:
\be
g=e^{\frac{\Phi}{2}}
\ee
In what follows
we will identify the renormalization group coordinate ${\mu}$ with a space-time
dimension. From the previous equations we derive the dependence on this 
coordinate of the dilaton field ${\Phi}$:
\be
{\Phi}({\mu})=-\log\log({\mu})+\log ({\b_0})
\ee
Next we use, in order to fix the string space-time geometry, the
sigma model beta function equations for the bosonic string:
\be
R_{{\mu}{\nu}}+{\nabla}_{\mu}{\nabla}_{\nu}{\Phi}=0
\ee
\be\label{escalar}
({\nabla}{\Phi})^{2} - {\nabla}^{2}{\Phi}  = C
\ee
With the constant $C$ given by:
\be
C= \frac{26-D}{3}
\ee\label{cte}
To solve these equations for the asymptotically free dilaton field we
will use the following ansatz for the metric:
\be
{ds}^{2}= a({\mu}){d\vec{x}}_d^{2} + b({\mu}){d{\mu}}^{2} + 
c({\mu}){d\vec{y}}_D^{2}
\ee
(where ${d\vec{x}}_n^{2}$ represents the ordinary Minkowski metric 
in $\mathbb{R}^n$).
Before motivating our ansatz for the metric let us just present the
solution. Introducing the variable $\rho$ by the relation:
\be
{\Phi}({\mu})= \log(\rho)
\ee
the metric solution is given by:
\be
a({\mu})=\rho
\ee
\be
b({\mu}){d{\mu}}^{2}={d{\rho}}^{2}
\ee
\be
c({\mu})=1
\ee
The corresponding metric is solution provided:
\begin{itemize}

\item The number of $x$ coordinates , $d=4$
\item The string is critical i.e the number of $y$ coordinates $D=21$.
and therefore the constant $C$ is equal zero.
\item The metric is a solution with both euclidean and 
Lorentzian signature.
\end{itemize}

Notice also that the sigma model beta functions we have used assume
already a vanishing vacuum expectation value for the closed string
tachyon background field.

\par

The metric we obtain as solution to the string beta function equations with
an asymptotically free behavior for the dilaton is of the type
first considered by Polyakov \cite{polyakov} with the function 
$a({\mu})$ playing the 
role of a running string tension. Moreover
we can interpret 
the variable ${\rho}$ as the analog of Liouville field
in non critical strings. It is important to
observe that the string beta function equations are not sensitive
to the 
constant in equation (\ref{escalar}) i.e they do not feel the concrete 
gauge group
we are working with. 

\par
Next we will briefly describe the different regions of our space-time. The 
region corresponding to ultraviolet behavior of the Yang Mills theory is
${\mu}= \infty$. In this region the scalar curvature that is given by:
\be
R= - \frac{1}{{\rho}^{2}}
\ee
is strong while in the strongly coupled field theory
infrared region we have weakly coupled gravity. At the point ${\rho}=0$ we
have a naked singularity that could be interpreted as the limit of validity
of the string description.
Comparing with Polyakov's type of metrics
we should notice that the singularity satisfy the zig-zag condition
$a({\mu}=1)=0$. Therefore it would be natural to define a string
representation of Wilson loops to impose Dirichlet boundary conditions
on ${\rho}=0$.
\par

As mentioned in the introduction the RGA to the string description
of non abelian gauge theories have as its goal to model in terms
of sigma model string beta function equations the different renormalization
group equations of the corresponding quantum field theory. In this sense it is
of major importance to have an interpretation of the string space-time metric
in purely quantum field theory terms. Here is where the role of Zamolodchikov's
$c$ theorem becomes important. As described in reference \cite{alvarezgomez}
 the $c$-theorem
can be understood as a direct consequence of gravitational focusing
for congruences of null geodesics in the {\em renormalization group} direction
${\mu}$. Therefore from the RGA point of view the curvature in the coordinate 
${\mu}$ should be intimately related with the evolution of $c$-function.
Here we will not enter into a detailed discussion on the definition
of the $c$-function from a geometrical point of view reducing ourselves
to discuss the phenomena of gravitational focusing for the asymptotically free
space-time metric.

\par
\subsection{Two loop approximation and universal metric}
Let us now consider the two loop running coupling constant for pure
Yang Mills (\cite{particle}):
\bea\label{tldilaton}
g^{2}({\mu})=\frac{16\pi^2}{\b_0\log\mu^2}&&
[ 1 - 
\frac{2\b_1}{\b_0^2}\frac{\log\log\mu^2}{\log\mu^2}
+ \frac{4\b_1^2}{\b_0^4 \log^2 \mu^2} \nonumber\\
&&((\log\log\mu^2 - 1/2)^2 + \frac{\b_2\b_0}{8\b_1^2} - \frac{5}{4})]
\eea
where the beta function is parametrized as:
\be
\b(g)\equiv \mu\frac{\pd g}{\pd \mu} = -\frac{\b_0}{(4\pi)^2} g^3 
-\frac{2\b_1}{(4\pi^2)^4} g^5 - \frac{\b_2}{2(4\pi)^6}g^7
\ee
The quotient $\b_1/\b_0^2 $ happens to be independent, for pure Yang Mills, 
of the particular gauge group
we are working with. (Up to two loops we should neglect $\b_2$).  Starting with
the three loop contribution,  a nontrivial dependence on the 
renormalization scheme appears.
The corresponding dilaton is given up to a constant by
\bea
\Phi(\m) = -\log\log{\mu^2}+ \log&[& 1 -\frac{2\b_1}{\b_0^2}
\frac{\log\log\mu^2}
{\log\mu^2}
+ \frac{4\b_1^2}{\b_0^4 \log^2 \mu^2} \nonumber\\
&&((\log\log\mu^2 - 1/2)^2 + \frac{\b_2\b_0}{8\b_1^2} - \frac{5}{4})]
\eea
Our ansatz for the confining string metric is now dictated by the one loop
solution described in the previous section, namely
\be\label{metric}
ds^2= a(\mu)d\vec{x}_d^2 + b(\m) d\mu^2 + d\vec{y}_D^2
\ee
It is a simple matter (with the help of some algebraic manipulations) 
to check that 
the metric:
\be\label{universal}
{ds}^{2}= e^{{\Phi}}{d\vec{x}}_d^{2} + (d e^{{\Phi}})^{2} + 
{d\vec{y}}_D^{2}
\ee
with ${\Phi}$ the one given by the quantum field theory renormalization group
equation, is the desired solution to the sigma model beta function equations.
It is important to
stress that here also ( as already did happen in  the one and two loop 
cases) the sigma model
beta function for the dilaton $\b_{\Phi}=0$ implies that the metric 
(\ref{metric}) is solution only when we consider a four dimensional space-time
and a critical string i.e $D$ in (\ref{metric}) equal $21$. 
\par
At this point the reader can wonder why - if we are identifying the
string coupling constant with the Yang Mills coupling- we are 
using ,in order to match Yang Mills
higher loop effects,  the  sigma model beta function equations at string
tree level. This is a very deep question whose answer could be related to the 
fact that dilaton tadpole effects of Fischler-Susskind type 
{\cite{susskind}} are absorbed
in renormalizations of the metric that preserve the universal form 
(\ref{universal}).

\par
The result we have presented for the two loop case is not accidental,
in fact we can prove the following theorem
\par
\textbf{Theorem}. The universal metric (\ref{universal}) is  
(for an arbitrary dilaton field depending only on the renormalization
group variable) an exact solution to the critical string 
sigma model beta 
function equations. 
\par
This means that {\em exactly} the same spacetime metric conveys quite
different solutions, summarizing the whole perturbative expansion
of the gauge theory. A lot of physics is hidden in the change
of variables needed to put the metric in the universal form (\ref{universal}).

This is by itself an interesting property of
the tree level sigma model beta function equations that depends
crucially on working at criticality i.e for the constant (\ref{cte}) 
vanishing.

\subsection{Generalities on the Confining Manifold}
In terms of the universal metric (\ref{universal})
it is convenient for most geometrical studies to employ
the variable
\be
\rho\equiv e^{{\Phi}}
\ee
such that the metric reads
\be
ds^2 = \rho d\vec{x}_4^2 + d\rho^2
\ee
and the dilaton
\be
\Phi = \log \rho
\ee
The natural range for $\rho$ is $(0,\infty)$ due to the singularity
at $\rho=0$.
It is amusing
to note that we can extend the metric to the range $(-\infty,0)$.
This is a non-analytic extension, in the sense that the complete metric 
would read:
\be
ds^2 = |\rho| d\vec{x}_4^2 + d\rho^2
\ee
and the dilaton
\be
\Phi = \log |\rho|
\ee
This non analytic extension of the metric could be related to the existence
of a deconfining phase transition.
\par
We shall use roman capitals to denote generic five-dimensional indices,
 $A,B,C,\ldots =0,1,2,3,4$, where $x^4\equiv z$; and greek letters for the
ordinary four-dimensional variables, i.e.
$\m ,\n, \ldots = 0,1,2,3 $, whereas $i,j,k,\ldots = 1,2,3$.
\par
The isometry group of the confining manifold is just the four-dimensional
Poincar\'e group, $IO(1,3)$. The four Killings associated to four-dimensional
translations are:
\be
k_{\m}\equiv \pd_{\m}
\ee
Note in particular that the {\em contravariant} components of the Killing
associated to time translations are
\be
k^A\equiv (1,0,0,0,0)
\ee
%
\subsection{Gravitational Focusing}
The first ingredient we need are the null geodesics for our metric. They are 
given in affine parametrization by:
\be
x^{\m}=\frac{2 c^{\m}\gamma}{c^0}
(\frac{3c^0}{2\gamma}(\tau - \tau_0) + \rho_0^{3/2})^{1/3} + x_0^{\m} - 
\frac{2c^{\m}\gamma\rho_0^{1/2}}{c^0}
\ee
and
\be
\rho = (\frac{3c^0}{2\gamma}(\tau - \tau_0 + \rho_0^{3/2})^{2/3}
\ee
with $\eta_{\m\n}c^{\m}c^{\n} = - \frac{c^0}{\gamma}^2$. 
\par
The covariant null tangent vector is
\be
k_{\m}=c_{\m}
\ee
(where $c_{\m}\equiv \eta_{\m\n}c^{\n}$) and
\be
k_4=\frac{c^0}{\gamma} \rho^{-1/2} = 
\frac{c^0}{\gamma} (\frac{3c^0}{2\gamma}(\tau - \tau_0) + \rho_0^{3/2})^{- 1/3}
\ee
The affine parameter is normalized in such a way that
\be
k^A \nabla_A \tau = 1
\ee

The covariant derivative is given by:
\bea
\nabla_{\m}k_{\n}=\frac{1}{4}k_4 \eta_{\m\n}\nonumber\\
\nabla_4 k_{\m} =\nabla_{\m} k_4 = - \frac{1}{2\rho}k_{\m}\nonumber\\
\nabla_4 k_4 = - \frac{c^0}{2\gamma\rho^{3/2}}
\eea
\par

The fact that the antisymmetric part vanishes is equivalent to the fact that 
the
vorticity of the corresponding congruence $\omega = 0$. This means that
the congruence is hypersurface orthogonal.
\par
The expansion is given by
\be
\theta = \frac{3c^0}{2\gamma\rho^{3/2}}
\ee
and the shear by
\be
\sigma^2 = - \frac{15 c^0}{8 \gamma^2\rho^3}
\ee 
\par
We can see directly a c-theorem of sorts operating in the confining metric:
the expansion $\theta$ behaves in physical variables as
\be
\theta\sim (\log \m )^{3/2}
\ee
being ${\infty }$ at the singularity as it should be.
This means that the corresponding c-function
as defined in \cite{akhmedov}  will also have a similar physical
behavior.
This definition, however was tailored for  holographic theories, and
it is conceivable that some modifications are needed in our case.
It is curious ,however, to notice how easy is to prove this sort
of properties in the string
representation  in 
contradistinction to the painful attempts to prove it directly by quantum
field theory methods \cite{latorre}. 
Clearly further research is needed on what 
quantum field theoretical correlators is the c-function measuring.

\par

\subsection{Other geodesics}
In order to obtain analytic expressions, it is useful to parametrize
the curves in terms of the time coordinate, $t$.
The Minkowskian space-like coordinates have always a form identical to the one
corresponding to flat space geodesics, namely:
\be
x^i = x_0^i + \beta^i t
\ee
(where $\beta^i\equiv\frac{c^i}{c^0}$). Let us now examine 
several subcases in turn.
\subsubsection{Null geodesics}
Null geodesics exist only when $\vec{\beta}^2 \equiv\beta^2 \leq 1$ 
(that is, when the 
Minkowskian projection is timelike). Its explicit form in our present 
parametrization is:
\be
\rho = \frac{1-\beta^2}{4}(t - t_0)^2
\ee
when $\vec{\beta}^2 < 1$, and
\be
\rho = \rho_0
\ee
when $\vec{\beta}^2 = 1$.

\subsubsection{Timelike geodesics}
Timelike geodesics also need $\vec{\beta}^2 < 1$. The explicit expression
is
\be
\rho = \frac{(c^0)^2}{2}(1-\beta^2)(1 + \sin\frac{t - t_0}
{c^0 \sqrt{1-\beta^2}})
\ee
It is remarkable that this implies the existence of {\em closed
timelike geodesics}; for example, when $\beta^i = 0$:
\bea
x^i &=& x^i_0\nonumber\\
\rho &=& \frac{(c^0)^2}{2\gamma^2}(1+\sin\frac{\gamma(t-t_0)}{c^0})
\eea
The physical meaning of five-dimensional timelike geodesics is related to
unsuspected properties of the renormalization group flow, worth examining
in some detail. We stress once more that the  ordinary four dimensional
projection never suffers from closed timelike curves; they close in the
coordinate associated to the renormalization group evolution parameter only. 
\subsubsection{Spacelike geodesics}
Spacelike geodesics appear for any value of $\beta$. When $\beta\neq 0$,
\be
\rho = - (c^0)^2 (1 - \beta^2) \cosh^2 \frac{t - t_0}{c^0 \sqrt{1-\beta^2}}
\ee
and when $\beta = 1$,
\be
\rho = \rho_0 \pm t
\ee

\section{Energy functional}
Since we are considering a background for the critical bosonic string
an issue we should address is the fate of tachyons on this particular
space-time. Here we will follow the approach of \cite{breitenlohner} extended
from the case of Anti-de-Sitter (AdS) to our confining metric. 
Let us first recall the 
AdS argument.
 We consider the case of a free scalar field of mass $m^{2}$. Next
conservation of the energy momentum tensor is enforced. This implies certain
boundary conditions on the the fields at the boundary. Finally we solve
the field equations on the functional space satisfaying these boundary 
conditions. For stationary wave solutions this amounts to a bound on
allowed values for the mass $m^{2}$. Certain negative values of $m^{2}$ are
allowed by this bound corresponding to tachyons with Compton wavelength of the
order of AdS curvature radius. However fluctuations corresponding to these
tachyons have positive total energy impliying the stability of AdS space-time.
In the AdS case the Cauchy problem  on the reduced functional space, 
determined
by the condition of energy conservation, has a unique solution.
In order to extend the previous analysis to our case the first thing we
will do is to derive the boundary conditions on the wave functions required
by conservation of energy.
\par
Given an arbitrary scalar field defined on the confining manifold with
energy-momentum tensor $T_{AB}$, there is a {\em Killing energy} current,
namely
\be
j^{A}\equiv T^{AB} k_{B}
\ee
which is covariantly conserved, 
\footnote{ Please note that we need at least two scalar fields to build
a conserved {\em number} current, namely
\be
N^A \equiv e^{\Phi}(\phi_2\pd^A \phi_1-\phi_1\pd^A \phi_2)
\ee
}
i.e.
\be\label{div}
\nabla_{A}j^{A} = 0
\ee
Applying Stokes'theorem to the integral over the five-dimensional 
region $\mathcal{R}$ delimited by
the hypersurfaces $t = t_0$ and $t = t_1$, of the five-form proportional
to the first member of (\ref{div}), i.e. of $d* j$, where $j$ is 
the one-form dual to
the current, $j\equiv j_{A}dx^{A}$, we get
\be
0 = \int_{\mathcal{M}} d*j = \int_{\pd \mathcal{M}} *j = E(t_2)-E(t_1) +
 \Delta E
\ee
where
\be
E(t)\equiv \int d^3 x dz j^0(t,x^i,z)
\ee
and
the flux over the boundary at infinity
\be
\Delta E = \lim_{(x^i,z)\rightarrow \infty} \int [(\epsilon_{ijk}dt\wedge dx^j
\wedge dx^k
\wedge dz) j^i + d^4x j^4]
\ee
Only when the physical boundary conditions are such that 
\be
\Delta E =0
\ee
there is actual energy conservation,
\be
\frac{dE}{dt}=0
\ee
For a scalar field of mass $m$ it is readily found that
\be
T_{AB} = \frac{1}{2}e^{\Phi}[\nabla_A \phi\nabla_B \phi - \frac{1}{2}g_{AB}
(g^{CD}\nabla_C \phi\nabla_D \phi - m^2 \phi^2)]
\ee
so that the five dimensional energy is:
\be
E[\phi]\equiv\frac{1}{4}\int d^3x dz 
[\dot{\phi}^2 + (\nabla \phi)^2 +z( m^2 \phi^2 +(\phi')^2]
\ee
In the {\em free} case the energy is zero on shell, provided the fields
vanish at infinity fast enough, and this is true no matter what the sign 
of $m^2$ is, i.e., even for the putative tachyon.
\subsection{Variations of the energy functional}
Because we are mostly interested in the stability of our solution we
will reduce ourselves to study the spectral problem for the operator
defined by the second variation of the energy around $T=0$ 
corresponding to vanishing tachyonic background. We will use static 
fluctuations. Notice that the first static variation of the energy is zero
since we are working with a solution of the equations of motion. 
\par
The second
variation
can actually be written in our case as:
\be
\frac{\delta^2 E[\phi]}{\delta \phi^2}= 2 \int d^3 x dz[-\Delta_3 + L]
\ee
where $L$ is the operator:
\be
L= - z\frac{d^2}{dz^2}-\frac{d}{dz} + z m^2
\ee
The issue as to the stability of the solution is then equivalent to
the issue of whether the operator $L$ has any negative eigenvalues
(on the functional space in which all conservation laws can be actually
implemented).We are then led to study the eigenvalue problem
\be\label{taq}
L T = \lambda T
\ee
This is a second order ordinary differential equation, with a regular
singular point at the origin and an irregular one at infinity.
By studying the inditial equation it is inmediatly obvious that one
of the solutions has a logarithmic singularity at the origin.
\par
In order to be able to study simultaneously the two cases of positive
and negative mass square we shall for the latter use $m^2 \equiv -\mu^2$.
By performing the change of both dependent and independent variables:
\be
T(z) = e^{-z\xi} y(z/\rho)
\ee
(where $\xi^2=m^2$ and $2\xi\rho=1$), the equation is reduced to
a confluent hypergeometric one:
\be
\frac{d^2 y}{dx^2} + (\frac{1}{x} - 1) \frac{dy}{dx} + 
(\frac{\lambda}{2m}-\frac{1}{2})\frac{1}{x} y = 0
\ee 
Let is define $a\equiv \frac{m-\lambda}{2m}$.
There is an exceptional case, when $a\in - \mathbb{N}$
(that is, $\lambda = (2p+1)m$, with $p\in \mathbb{Z}^{+}$),
where the solution is given in terms of Laguerre polynomials,
\be
y = L^{(0)}_n (x)
\ee
Otherwise the two independent solutions are given in terms of the 
functions (\cite{erdelyi}) $ \phi(a,1,x)$ and $\psi(a,1,x)$,
\par
The function $\phi(a,1,x)$ is regular at the origin
\be
\phi(a,1,x)\equiv ~_{1}F_{1}(a,1,x)\sim_{x\sim 0} 1 + a x + \ldots
\ee
and at infinity behaves as
\be
\phi(a,1,x)\sim_{x\sim\infty} \frac{1}{\Gamma(a)}e^x x^{a-1}
\ee
On the other hand the function $\psi(a,1,x)$ has got a logarithmic
divergence at the origin
\be
\psi(a,1,x)\sim_{x\sim 0} -\frac{1}{\Gamma(a)} \log[x] + \ldots
\ee
whereas at infinity behaves as
\be
\phi(a,1,x)\sim_{x\sim\infty}  x^{- a}
\ee
When $m^2 >0$ the  only solution regular at the origin is fine. 
We can arrange our variables
in such a way that asymptotically
\be
T(z) \sim_{x\sim\infty} e^{- m z} (2mz)^{a-1}
\ee
This means that the solutions and their derivatives are square integrable
on $dz$.
\par
When $m^2 <0$, however, the asymptotic behaviour of the putative solution
is
\be
T\sim e^{\pm i\mu z}(\pm 2i\mu z)^{a-1}
\ee
(where $a = \frac{1}{2} + \frac{i\lambda}{2\mu}$), too slow to be convergent,
for any value of $\lambda$.
\par
Thus we can conclude that there is not any negative eigenvalue -even
in the tachyonic case $m^{2}<0$-  with normalizable wave function. In other
words on the functional space defined by the boundary conditions required
by energy conservation the operator $L$ has not any negative eigenvalue , which
we interpret as a strong indication of the stability of our confining metric.
Comparing with the bound in \cite{breitenlohner} we observe that for
normalizable static wave functions the mass $m^{2}$ is forced to be $\ge0$. The
extension of the previous anlysis to stationary wave functions does not change
the result.
\par
It is is illustrative to present the previous analysis
from the point of view of an equivalent Schrodinger problem. In order to do 
that let us  
perform the change of variables in equation (\ref{taq})
\be
T(z)= t(z)z^{-\frac{1}{2}}
\ee
and the equation reduces to
\be
\frac{d^2 t}{dz^2} + [ \frac{\lambda}{z} + \frac{1}{4 z^2} - m^2]t = 0,
\ee
an equation first studied by Whittaker, which is an Schr\"odinger
equation for the reduced potential
\be
U(z)= -\frac{1 + 4\lambda z}{4 z^2}
\ee
and reduced energy
\be
\epsilon = - m^2
\ee
Please note that now we are viewing the problem head down: we are assuming
we already know $\lambda$, and we look for the energy as an eigenvalue.
The potential starts at $U_{-} = -\infty $ at $z=0$ can become positive if
$\lambda$ is negative enough,and reaches $U_{+}=0$ at $z=\infty$, but {\em not
faster} than $1/z$ (which is a necessary condition for most inverse stattering 
theorems to hold)
\par
Again, for $m^2>0$, we need the potential to have a negative eigenvalue,
which is no problem (there is indeed a continuum spectrum 
$\epsilon\in(0,-\infty)$). 
\par
For $m^2<0$, we would need a positive energy ,
which is not possible with normalizable states, owing to the slow asymptotic
behaviour of the potential at infinity.
\par
The massless case is marginal.

\section{The propagation of scalar disturbances}

From a practical point of view the main utility of a confining 
string representation of a non-abelian gauge theory ought to be the computation
of the glueball spectrum (which is in principle measurable, although the
widths are quite big, and the mixing with $q\bar{q}$ states is important).
\par
The main tool in this computation is a calculation of two Wilson loops
correlator,
\be
<W(C)W(C')>
\ee
which conveys the desired information in the location of its poles.
While we certainly plan to perform this computation in the near future,
we would like here to examine a very preliminary aspect of this problem,
namely the semiclassical propagation of a scalar disturbance in the
confining metric (which could be a crude model of the $0^{++}$ glueball).
 
We shall consider fields in the four-dimensional momentum representation,
i.e.
\be
\phi(x,z)\equiv\int d^4 k e^{i k.x} \phi(k,z)
\ee
The wave equation for massless (in the five dimensional sense)
scalar disturbances is then
\be
z\frac{d^2 \phi}{d z^2} + \frac{d\phi}{dz} = k^2 \phi
\ee
It is plain to check that there are no normalizable solutions
when $k^2 = 0$ (which corresponds to the massless case from
the four-dimensional point of view).Actually, the general solution is then
\be
\phi(k^2 =0,z) = c_1 \log z + c_2
\ee
This is enough to stablish the existence of a {\em mass gap}, in the sense that
massless disturbances from the five-dimensional sense neccessarily correspond
to massive ones from the four-dimensional one.
\par
When $k^2\neq 0$ the solution is easily expressed in terms of Bessel functions;
\be
\phi(k,z) = B_0(2 k \sqrt{z})
\ee
where $B_0(x)$ represents a Bessel function; actually a linear combination
of the functions $J_0(x)$ and $Y_0(x)$ when $k^2<0$ (in this case, the argument
in the function should be $2\sqrt{-k^2} \sqrt{z}$) , and a linear combination
of the modified functions $I_0(x)$ and $K_0(x)$ when $k^2 >0$.
\par
In the former case there are no normalizable solutions (in accordance with
the results in the last paragraph). Physically, this means that there are
no {\em four-dimensional tachyons} propagating. In the latter, 
normalizability again 
uniquely selects the $K_0$ solution.

\par
Normalization now means, for example, that
\be
\int_0^{\infty} \phi(z)^2 dz =1
\ee
But this integral is easily shown to be equivalent to
\be
\int_0^{\infty} K_0^2 z dz = 2 k^2 = \frac{1}{2}
\ee
enforcing a discrete {\em four dimensional} mass spectrum, namely 
$m_4^2 = \frac{1}{4}$. Energy 
is of course measured in units of $\Lambda$, so that this gives 
\be
m= \frac{\Lambda}{2}.
\ee
This value seems a bit low, \footnote{ Experimentally the situation is
unclear; although in \cite{particle} several candidates are reported, 
notably the $f_0(1500)$. In \cite{forshaw} evidence is reported for a Pomeron
trajectory $\alpha_P(t)=\a_P(0) + \a'_P t$, with $\a_P(0)= 1.08 GeV$
and $\a'_P = 0.25 GeV^{-2}$. The stardard lattice value (which seems to
 be a decreasing function of the $N$ factor in the gauge group $SU(N)$), is
about $m\sim 3.64 \sqrt{\sigma}\sim 2.18 GeV$ \cite{teper}. In order
to compare with the string value, an extrapolation towards $N=\infty$ ought to
be made.}but of course this is only but a crude estimate and we do not 
want to push it too far.

\section{The Wilson loop and the soft dilaton theorem}

An important test of any confining string model is to produce an
area law behavior for the vacuum expectation value of the static Wilson loop,
which we shall take in the static gauge
\bea
x^0 = \tau\nonumber\\
x^1=\sigma 
\eea
and the world sheet of the confining string will be characterized by
the time translation of the profile
\be
\rho=\rho(\sigma)
\ee
with $\rho(\pm l/2)=0$.
Once we
have a string model of a pure gauge theory we should represent the 
vacuum expectation value for the Wilson loop as a sum over random surfaces
with fixed boundary as defined by the loop. In the semiclassical approximation
we can use the corresponding Nambu-Goto form of the string action and to
look for the minimal area surface bounded by the loop {\cite{wilsonmalda}}. 
\par
For the confining string metric the Nambu Goto action is given by:
\be
S=\frac{1}{2\pi\alpha'}\int\sqrt{\rho((\partial_{x}\rho)^{2}+\rho)}
\ee
where by simplicity we use a static loop with profile given
in terms of the function $\rho(x)$. Denoting $\rho_{0}$ the minimun
value of $\rho(x)$ at the symmetric point $x=0$ it is easy to see that
this valu depends on $L$ ,the spatial length of the loop, as 
$\rho_{0}\thicksim L^{2}$.
\par
If we use in the Nambu Goto action a constant value for the string
tension $\alpha'$ the potential energy in this semiclassical approximation
will go as:
\be
V(L)\thicksim L^{3}
\ee
which means overconfining.It is important to stress that this result is
generic for our metric independently if we locate the loop at
$\rho=0$ or at $\rho=infty$. The only difference is that
when located in the zig-zag hyperplane $\rho=0$ the minimal surface
will live in the non analytic extension of the space corresponding to
negative values of $\rho$ ( i.e to values of $\mu$ between zero and one ).
\par
An important theorem for the bosonic string is the so called soft dilaton 
theorem that stablish the following interesting relation between
the string coupling constant and the string tension:
\be
lim_{k\to 0}A(k;p_{1},p_{2},...p_{n})=
cg(\sqrt\alpha')^{\frac{(d-2)}{2}}(\sqrt\alpha'\frac{\partial}
{\partial{\sqrt\alpha'}}-\frac{1}{2}(d-2)g\frac{\partial}{\partial{g}})
A(p_{1},p_{2},...p_{n})
\ee
Where $A(k;p_{1},p_{2},...p_{n})$ is the amplitude for a soft
dilaton insertion of momentum $k$ and $n$ external gluon vertex operators
with momentum $(p_{1},p_{2},...p_{n})$. If now we identify the open
string coupling constant g with the Yang Mills coupling, something
that we have done along our construction them the soft dilaton theorem
{\cite{ademolo}}
will implies that if g is running with the renormalization group parameter
$\mu$ them $\sqrt{\alpha'}^{\frac{(d-2)}{2}}$ should change with $\mu$ in
precisely the opposite way. Notice also that in the soft dilaton theorem
expression we must consider the momentum $(p_{1},p_{2},...p_{n})$ living in 
the four dimensional physical space-time. This is necessary in order to be
consistent with our identification of the string coupling constant
with the gauge coupling thus we will take $d=4$.
\par
The way we will use now the soft dilaton theorem is by simply changing
the string tension in the Nambu-Goto action by the effective running
string tension dictated by the soft dilaton theorem relation between 
the scaling of g and $\alpha'$. In these conditions we get:
\be
S=\frac{1}{2\pi}\int\sqrt{\rho^{-1} ((\partial_{x}\rho)^{2}+\rho)}
\ee
It is now plain to see that the potential energy in this case goes
like
\be
V(L)\thicksim L
\ee
i.e as the area law.
The reader can wonder what will happen if instead of taking $d=4$
in the soft dilaton theorem we leave free this value. The potential we
will get is given in terms of $d$  as
\be
V(L)\thicksim L^{5-d}
\ee
producing the area law only in the four dimensional case.
\par
The reader should notice that the Wilson loop is ,properly speaking, the first 
string computation we are performing. Before
we were considering semiclassical gravity effects 
associated with our confining string metric. Very likely the way the soft
dilaton theorem enters into de computation of the Wilson loop is
related to loop equations understood as renormalization
group equations for the vacuum expectation value of the Wilson loop.
It is also important to stress that the soft dilaton theorem enssures
the area law behavior for the universal confining metric (\ref{universal})
whatever the loop order of the Yang Mills beta function we choose to
fix the dilaton running.
\par
There is still a problem we have not touched along our paper, that
can be relevant for our discussion of the Wilson loop computation, namely
the dynamical role of the $21$ spactator extra coordinates we need to add 
in order to find solution to the sigma model beta function equations. We
have not a definitive answer to this issue although we feel very meaningful
the fact that  our solution make it  compelling to work in 
a critical string framework.
( incidentally notice that this would be also the case even 
in the type $0$ strings setup with world sheet supersymmetry ). A potential
way to interpret these extra spectators is to work in non critical dimension
with an appropiate vacuum expectation value for the tachyon. In this case, and
in order to solve the tachyon beta function equation we need to invoke
some mechanism of the type described in reference (\cite{klebanov}) for the 
generation
of a tachyon potential $V(T)$ such that $mT_{0} +V'(T_{0}) =0$
where $T_{0}$ is the required tachyon vacuum expectation value. In spite
of the fact that this can be an interesting mechanism to avoid the presence
of extra spectators we insist,
until reaching a deeper understanding on
the meaning of criticality, on the beauty of having an exact string solution
with a running dilaton consistent with higher order effects in
quantum field perturbation theory.

\section*{Acknowledgments}
We are grateful for stimulating discussions with Luis Alvarez-Gaum\'e,
Peter Hasenfratz and Peter Minkowski. We are indebted to Gabriele Veneziano
for pointing out an error in an earlier version of the manuscript.
This work ~~has been partially supported by the
European Union TMR program FMRX-CT96-0012 {\sl Integrability,
  Non-perturbative Effects, and Symmetry in Quantum Field Theory} and
by the Spanish grant AEN96-1655.  The work of E.A.~has also been
supported by the European Union TMR program ERBFMRX-CT96-0090 {\sl 
Beyond the Standard model} 
 and  the Spanish grant  AEN96-1664.



\end{document}